# Controlling the thickness of Josephson tunnel barriers with atomic layer deposition

Alan J. Elliot, Chunrui Ma, Rongtao Lu, Melisa Xin, Siyuan Han, Judy Z. Wu, Ridwan Sakidja, Haifeng Yu

*Abstract*— **Atomic Layer Deposition (ALD) is a promising technique for producing Josephson junctions (JJs) with lower defect densities for qubit applications. A key problem with using ALD for JJs is the interfacial layer (IL) that develops underneath the tunnel barrier. An IL up to 2 nm forms between ALD $Al_2O_3$ and Al. However, the IL thickness is unknown for ALD films < 1 nm. In this work, Nb-Al/ALD-$Al_2O_3$/Nb trilayers with tunnel barriers from 0.6 – 1.6 nm were grown *in situ*. Nb-Al/$AlO_x$/Nb JJs with thermally oxidized tunnel barrier were produced for reference. $R_N$ was obtained using a four-point method at 300 K. $J_C$, and its dependence on barrier thickness, was calculated from the Ambegaokar-Baratoff formula. The Al surface was modeled using *ab initio* molecular dynamics to study the nucleation of $Al_2O_3$ on Al. Current voltage characteristics were taken at 4 K to corroborate the room temperature measurements. Together, these results suggest that ALD may be used to grow an ultrathin, uniform tunnel barrier with controllable tunnel resistance and $J_C$, but a thin IL develops during the nucleation stage of ALD growth that may disqualify Al as a suitable wetting layer for ALD JJ based qubits.**

*Index Terms*— ***Ab-initio*** **molecular dynamics, atomic layer deposition, *in situ* deposition, Josephson junctions, quantum computing.**

## I. Introduction

The Josephson Junction (JJ), a superconductor-insulator-superconductor (SIS) tunnel junction, is a strong candidate for the implementation of quantum bits (qubits) due to its compatibility with modern semiconductor processing technology. This tunnel barrier (I layer) must be ultrathin (~1 nm) in order to maintain phase coherence across the superconducting electrodes, and because the critical current through the JJ decays exponentially with tunnel barrier thickness [1]. Producing an ultrathin, uniform, and leak-free tunnel barrier is difficult on metal substrates due to the naturally formed native oxides on most metals. However, traditional JJ fabrication techniques have cleverly used the easy of oxidation of Al to produce the tunnel barrier by thermally oxidizing a thin (~5 nm) "wetting layer" of Al to form Nb-Al/$AlO_x$/Nb JJs. This thermal oxidation scheme has been used to create high quality JJs for commercial applications such as SQUIDs and voltage standards. However, when JJs are employed for qubits, a more stringent requirement for lower noise arises to avoid superconducting phase decoherence, which makes meaningful quantum computation impossible. One major source of noise is two-level fluctuators (TLFs) in the insulating films of the qubit circuit [2]. Defects in the tunnel barrier arise from the process of thermally oxidizing aluminum and are a primary cause of TLFs [3]. While efforts to shield JJ qubit circuits from environmental noise to increase coherence times have met success [4], the root cause of the decoherence, and thus its ultimate solution, lies in the insulating materials used to in the circuit, primarily the tunnel barrier. Because thermal oxidation is the primary source of defects in the tunnel barrier, it is imperative to find novel methods of ultrathin tunnel barrier fabrication.

Atomic Layer Deposition (ALD) is a promising and unexplored alternative to thermal oxidation. Depicted schematically in **Fig. 1**, ALD is a chemical vapor growth method that uses self-limited surface reactions to grow films one molecular layer at a time, yielding subnanometer thickness control [5, 6]. Using $Al_2O_3$ as an example, alternating pulses of $H_2O$ and trimethylaluminum (TMA) are brought to substrates heated to ~200 °C by an inert carrier gas, usually $N_2$. The pulses are separated by a flush of $N_2$ to assure the two chemicals never meet in the gaseous state. In this way, the following reactions take place cyclically

$$AlOH^* + Al(CH_3)_3 \rightarrow AlOAl(CH_3)_2^* + CH_4 \quad (1)$$
$$AlCH_3^* + H_2O \rightarrow AlOH^* + CH_4 \quad (2)$$

where asterisks denote surface species. On some substrates, such as $SiO_2$, the surface is naturally hydroxylated **(Fig. 1a)**. Otherwise, the sample may become hydroxylated upon the first pulse of $H_2O$ **(Fig. 1b)**. The chamber is then flushed with $N_2$, and TMA is introduced **(Fig 1c)**. Ligand exchange occurs between the surface hydroxyl groups and the gaseous TMA. Because TMA only reacts with the surface hydroxyl groups and not itself, only one layer of $Al(CH_3)_2$ is deposited. Another flush occurs, and $H_2O$ is reintroduced **(Fig 1d)**. One molecular layer, or 1.2 Å, of $Al_2O_3$ is grown per cycle.

Manuscript received August 12, 2014. This work was supported in part by ARO contract No. ARO-W911NF-12-1-0412, and NSF contracts Nos. NSF-DMR-1105986 and NSF EPSCoR-0903806, and matching support from the State of Kansas through Kansas Technology Enterprise Corporation.

A.J. Elliot, C. Ma, R. Lu, M. Xin, S. Han, and J.Z. Wu are with the Department of Physics and Astronomy at the University of Kansas, Lawrence, Kansas 66044. Phone: 785-864-4626; fax: 785-864-5262; e-mail: alane@ku.edu

R. Sakidja is with the Department of Physics at the University of Missouri Kansas City, Kansas City, MO 64110.

H. Yu is with the Institute of Physics, Chinese Academy of Science, Beijing 100190

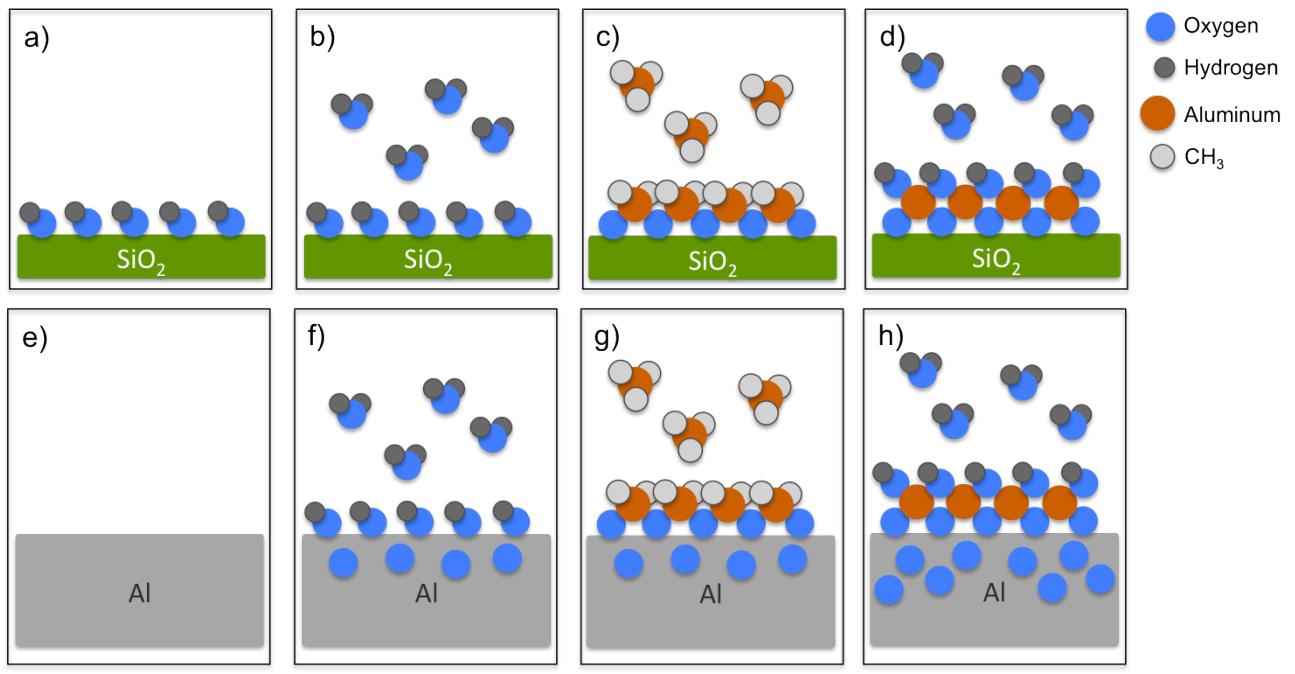

Fig. 1. Atomic Layer Deposition (ALD) $Al_2O_3$ growth occurs by pulsing $H_2O$ and TMA ($Al(CH_3)_3$) into the reaction chamber separately. In the case of SiO2, the surface is naturally hydroxylated, and the surface chemistry requirements for ALD are satisfied (a). First, $H_2O$ is introduced into the chamber to completely hydroxylate the surface of the substrate, (b). The $H_2O$ is evacuated from the chamber, then TMA is pulsed into the chamber to deposit one layer of Al (b). The TMA is then evacuated form the chamber, and $H_2O$ is reintroduced (d). On Al the necessary surface hydroxyl groups may not exist before the first $H_2O$ pulse (e). The first water pulse will both hydroxylate the surface and thermally oxidize a thin layer in the body of the film (f). When TMA is pulsed into the chamber, the reactions occur mainly on the surface (g). When water is reintroduced, further oxidation of the body of the Al film is possible if the combined aluminum oxide layer is not thick enough to act as a diffusion barrier (h).

ALD has several advantages over thermal oxidation for JJ tunnel barrier growth. First, while thermal oxidation creates an oxygen gradient from the surface down, the ALD reactions completely oxidize the surface to create a uniform film from the bottom up. Thus, ALD is expected to produce fewer point defects and TLFs. Also, thermal oxidation requires precise control over oxygen partial pressure and time to consistently produce the targeted critical current density ($J_C$) [1, 7]. In contrast, ALD's self-limiting growth mechanism allows substantial leeway in processing conditions while still yielding highly reproducible, subnanometer thickness control. Furthermore, while thermal oxidation limits material selection to $AlO_x$, many dielectric materials can be grown via ALD, including $Al_2O_3$, $HfO_2$, MgO, and ZnO (see reference [6] for a complete list). Some of these films, such as $HfO_2$, grow polycrystalline phases, which could be used to produce a single crystal tunnel barrier without laborious and expensive epitaxial techniques [6, 8]. These advantages make ALD a prime candidate for fabricating next generation JJ qubits with lower defect densities and higher coherence times.

However, the quality of ALD JJ tunnel barriers depends critically on the nucleation phase and the presence of interfacial layers (ILs). The chemical reactions of ALD (e.g. ALD $Al_2O_3$ given in eqns. 1-2) require the existence of surface species, such as OH* and $CH_3$*. This problem has been well studied on $SiO_2$, where the requirement is automatically satisfied by residual $H_2O$ on the surface **(Fig. 1a)**, and hydrogen terminated Si (H-Si) substrates, where surface activation is necessary to avoid a ~1 nm thick silicate interfacial layer [9, 10]. Similarly to $SiO_2$ and H-Si, metallic substrates have either reactive surfaces, such as Al and Cu, or inert surfaces, such as Pt and Au. In the latter case, nucleation of ALD films can be completely frustrated for the first 30-50 cycles of growth, which act as an incubation period to physisorb the ALD precursors onto the surface [8]. For reactive metals, such as Al, even *in situ* deposited films can acquire an IL up to ~2 nm thick when as little as 20 ALD cycles are performed [11, 12]. This IL is suspected to be caused by thermal oxidation of the Al substrate when it is exposed to $H_2O$ at 200 °C [11]. For example, in the ideal case of ALD occurring in a clean UHV chamber, the Al hydroxyl groups may not exist in abundance on the Al surface **(Fig. 1e)**. Upon the first exposure to $H_2O$ the surface will become hydroxylated, and oxygen will likely diffuse into the body of the Al film **(Fig. 1f)**. The penetration depth depends on the temperature of the sample, the pressure of oxygen, and the amount of time this oxygen pressure is maintained. When TMA is pulsed into the chamber **(Fig. 1g)**, it reacts only at the surface of the sample according to eqns. 1 and 2. Finally, when water is pulsed into the chamber again **(Fig. 1h)**, not only do the surface reactions in eqns. 1 and 2 take place, but more oxygen may diffuse into the Al film if the total oxide thickness is insufficient to act as a diffusion barrier.

*Ab-initio* molecular dynamics (AIMD) simulations were used to study the changes in Al surface chemistry that occur after an $H_2O$ pulse, and the results are given in **Fig. 2**. The AIMD simulations adopted the Bohn-Oppenheimer molecular dynamics implemented in VASP [13, 14] and used a 2x2 supercell of FCC Al under constant equilibrium volume and temperature (473 K). To simulate the initial Al surface, one $H_2O$ molecule was placed on the supercell to simulate the expected traces of $H_2O$ in the ALD chamber (**Fig. 2a**). After 6 ps of simulation time, no dissociation of $H_2O$ into OH was observed (**Fig 2b**). However, when multiple $H_2O$ molecules were placed on the surface to simulate an $H_2O$ pulse (**Fig. 2c**), dissociation occurred almost immediately. After only 3 ps of simulation time, the $H_2O$ molecules dissociated into $OH^-$ (**Fig. 2d**). We note the main mechanism of $OH^-$ formation during this initial period of AIMD simulation appears to be proton transfer between molecules, forming $H_3O^+$, which almost instantly dissociates into $H_2O_{ad}$ and $H^+_{ad}$. These simulations suggest the formation of $Al_2O_3$ becomes thermodynamically favorable under typical ALD processing conditions and that nucleation of ALD $Al_2O_3$ films is key and more importantly, can be controlled by the initial $H_2O$ exposure. Work to model the penetration of oxygen into the Al film is ongoing.

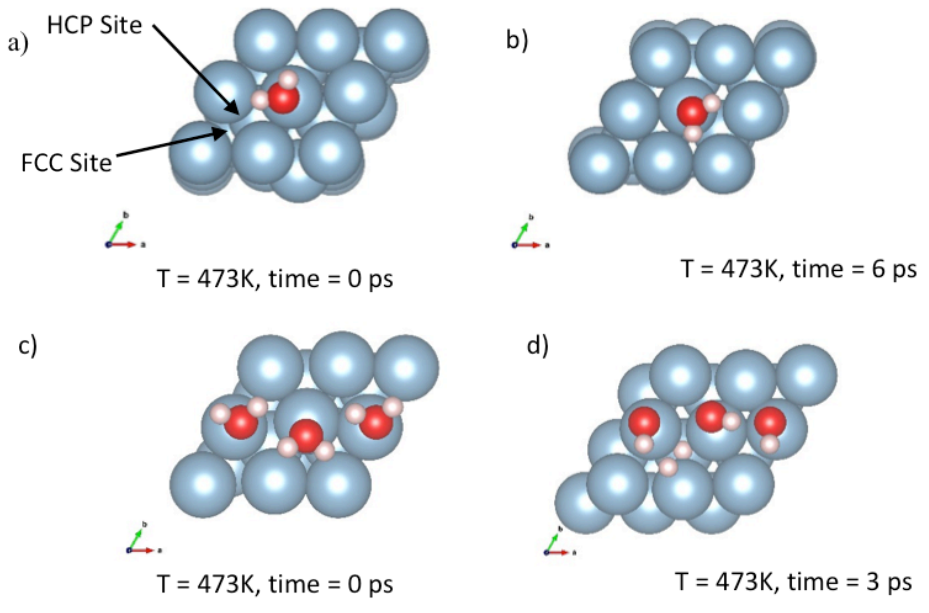

Fig. 2: Ab-Inito Molecular Dynamics (AIMD) simulations of $H_2O$ absorbing onto an Al surface. When only one $H_2O$ molecule is present on the Al surface, dissociation is thermodynamically unforable (a,b). However, when $H_2O$ molecules are in close contact with eachother on the surface, dissociation into OH and H is nearly instantaneous (c,d).

For JJ tunnel barriers, less than 10 ALD cycles are required for a thickness of ~1 nm. In this thickness regime, it is unknown whether $AlO_x$ or ALD-$Al_2O_3$ dominates the film. Furthermore, at the end of ALD growth the surface is populated with either hydroxyl or methyl groups, and our AIMD simulations suggest a high hydrogen mobility on the Al surface. The hydrogen in these surface groups may act as TLFs [15] or charge scattering centers [16], which could cause poor coherence in qubit applications or decrease the tunneling current through the JJ. The key challenges in the use of ALD for JJ tunnel barriers are to minimize the thickness of the $AlO_x$ IL and to produce a pristine $Al_2O_3$ top surface.

There are several possible ways to minimize the thickness of the IL and to clean the top surface of the ALD films, such as ion milling or processing condition optimization. However, we must first understand the influences on JJ performance before designing a solution. In this work, ALD was performed *in situ* to produce Nb-Al/ALD-$AL_2O_3$/Nb trilayers, which were then patterned into JJs. By varying the number of ALD cycles performed, extracting the projected $J_C$, and extrapolating the trend to 0 ALD cycles, we have found evidence that a thin $AlO_x$ IL forms even in this thickness regime. Low temperature measurements were used to corroborate these conclusions.

## II. Experimental

Nb-Al/ALD-$Al_2O_3$/Nb trilayers were fabricated in a homemade deposition system, which integrates UHV sputtering and ALD *in situ* [16, 17]. For comparison, traditional thermally oxidized Nb-Al/$AlO_x$/Nb trilayers were also fabricated with a target $J_C = 50$ A/cm$^2$. The Nb films were sputtered onto a water-cooled sample stage at 1.7 nm/s to minimize the formation of NbOx from trace oxygen. The bottom Nb was 150 nm, and the top Nb was 50 nm. The 7 nm Al wetting layers were sputtered at 0.5 nm/s. For trilayers with ALD tunnel barriers, 5 – 13 cycles (0.6 – 1.6 nm) of ALD-$Al_2O_3$ growth occurred at 200 °C with TMA and $H_2O$. For the trilayer with a thermally oxidized tunnel barrier, the Al wetting layer was exposed to 100 Torr $O_2$ for 3.5 hours.

The trilayers were patterned into JJ arrays using advanced lithography and etching techniques. Nominal JJ dimensions ranged from 7 – 10 µm, but profilometery was used to measure the actual dimensions, which were reduced by ~1.5 µm by etching. UV Photolithography was used to define the main wiring of the circuit, which includes four electrical leads for each of the 12 JJs on the circuit. The Nb films were etched using reactive ion etching in 15 mTorr $SF_6$ at 0.4 W/in$^2$. The $Al_2O_3$, $AlO_x$ and Al films were wet etched in 8% $H_3PO_4$. Electron Beam Lithography was used to define the junction area and the top wiring leads. The junctions were insulated by sputtered $SiO_2$. The tops of the junctions were cleaned in $Ar^+$ plasma before the top Nb wiring was sputtered.

The completed devices were characterized at room temperature and low temperature using a four-point configuration. At room temperature, current-voltage curves (IVCs) were taken using a semiconductor device analyzer (Agilent B5015A) with 25 µm tungsten probes. The room temperature resistances ($R_N$) were extracted from the slope of the IVCs. The samples were then cooled to 4 K using a physical properties measurement system (PPMS) from Quantum Designs. Current voltage characteristics were obtained using a custom, battery-operated measurement system independent of the PPMS.

## III. Results and Discussion

$R_N$ was measured at room temperature for samples with 5 – 13 ALD cycle tunnel barriers. $R_NA$, the specific tunnel resistance, was found to be independent of the JJ area, indicating a uniform tunnel barrier had been grown. **Fig. 3a** shows $R_NA$ vs. ALD cycles for all the trilayers in this study. The data point at 0 ALD cycles is from Reference [16]. It represents a sample that went through the heating and cooling process in the ALD chamber, but was not exposed to any ALD sources. Its deviation from the trend will be discussed presently. An exponential increase in $R_NA$ with increasing ALD cycles is observed, indicating $R_N$, and thus $J_C$, can be controlled with ALD.

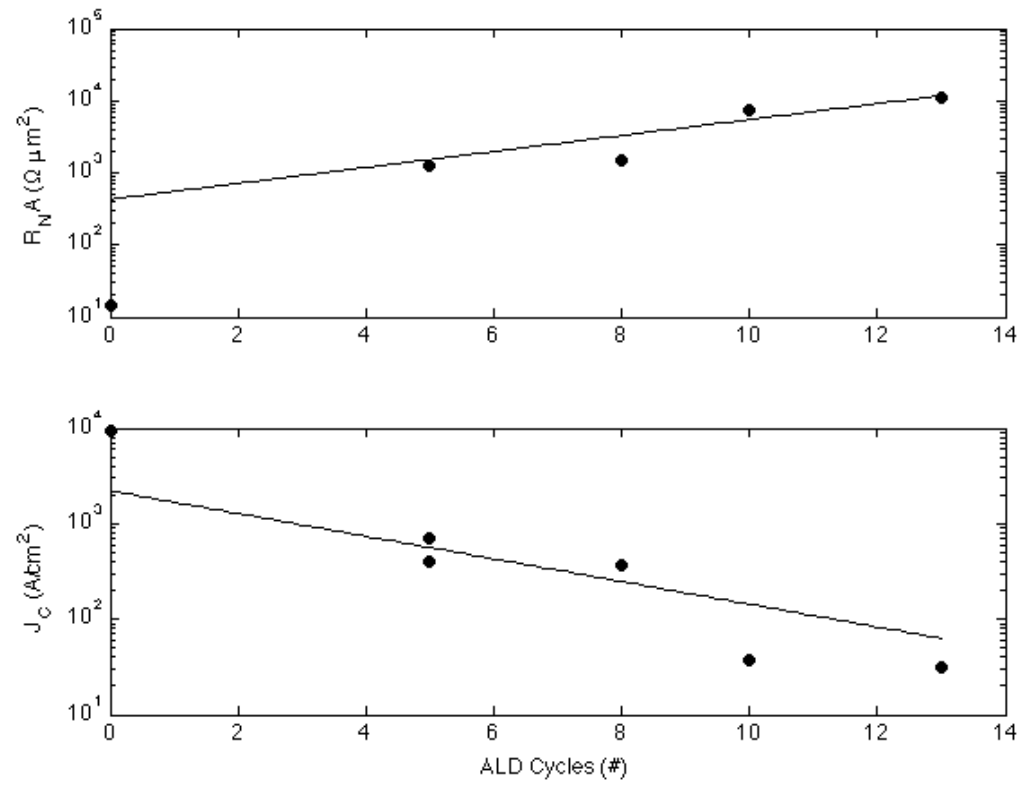

Fig. 3: Specific tunnel resistance vs ALD cycles (a), and projected $J_C$ vs ALD cycles (B). The data point at 0 is from Reference [16] and was not included in the fits.

By the Ambegaokar-Baratoff relation [18], $R_N$ is a function of $J_C$ and $A$

$$R_N = \frac{\pi\Delta}{2eJ_C}\frac{1}{A} \tag{4}$$

where $J_C$ is the critical current density, $R_N$ is the normal state resistance, $A$ is the area of the junction, $\Delta$ is the superconducting gap energy, and $e$ is the elementary charge. By fitting $R_N$ vs. $1/A$, $J_C$ can be extracted from the slope. **Fig. 3b** shows this calculated $J_C$ vs. ALD cycles. There is an exponential decrease in $J_C$ with increasing ALD cycles. The data point at 0 cycles was taken from Reference [16]. $J_C$ of this sample was measured at 4 K, in contrast to the other data points on the graph. The 0 cycle JJ was not included when fitting the trend to the data. At 0 cycles, the extrapolated $J_C \sim 1.1\text{x}10^3$ A/cm$^2$. According to the analysis in Reference [1], this $J_C$ value corresponds to a pinhole free Al surface and an oxidation dose of ~$10^5$ Pa-s. This indicates the formation of a thin $AlO_x$ IL. However, the extrapolated value does not agree well with $J_C \sim 9x10^3$ A/cm$^2$ measured on the 0 ALD cycle sample. This indicates an oxidation event occurs in the first few ALD cycles, which has been previously reported for ALD growth on W, Co, and Ta [19-21] and agrees with the AIMD simulations in **Fig. 2**. This oxidation event could also explain the deviation in **Fig. 3a**. Therefore, a thermally oxidized IL exists between the Al wetting layer and the ALD tunnel barrier that corresponds to a $J_C \sim 1.1x10^3$ A/cm$^2$. If this IL is sufficiently thick that $AlO_x$ dominates the tunnel barrier, pure Al wetting layers may be unsuitable for ALD growth of JJ tunnel barriers due to the presence of point defects in the $AlO_x$, and alternative wetting layers may be needed.

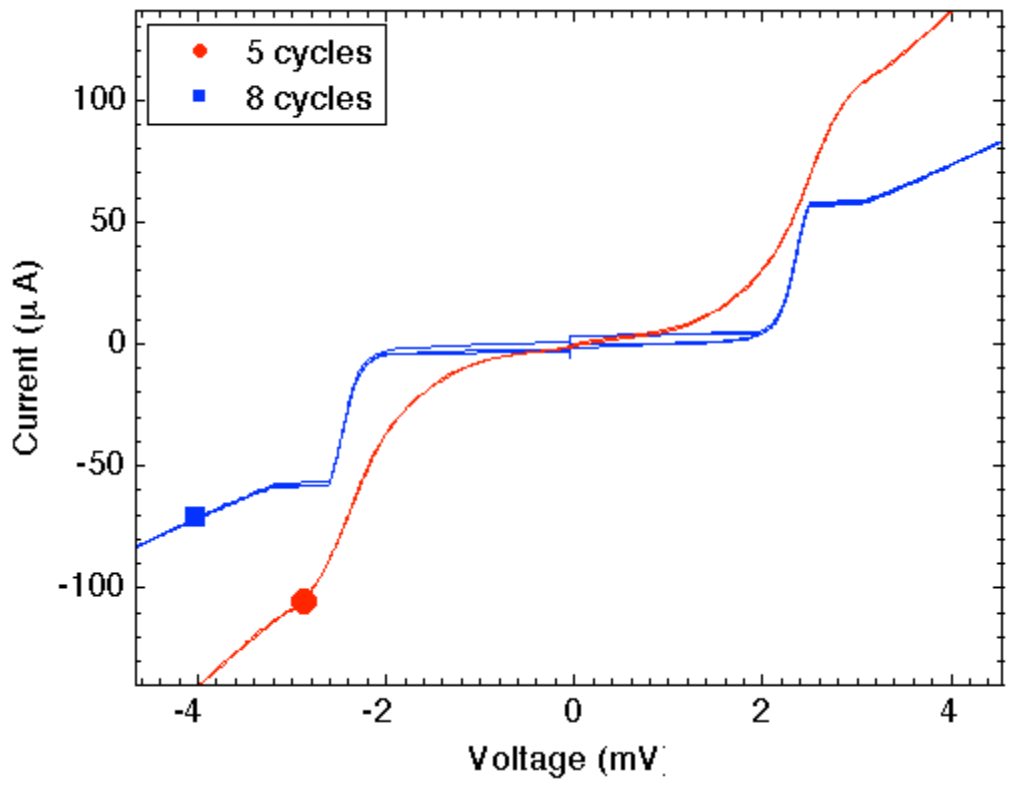

Fig. 4: Current Voltage Characteristics (IVCs) of two ALD JJs. The red (circle) IVC is from the present study and has a 5 ALD cycle tunnel barrier. There is significant distortion of the subgap region and the supercurrent is absent due to excessive noise in the measurement system. The blue (square) IVC has a 8 cycle tunnel barrier was patterned and measured in another lab [16].

While measurements of resistivity at room temperature provide important information about the quality of ALD $Al_2O_3$ tunnel barriers, and the results strongly indicate the formation of uniform, low-leakage tunnel barriers, a low temperature measurement of quasi-particle tunneling characteristics is the ultimate test to determine the integrity of the tunnel barrier. Current voltage characteristics (IVCs) of 10x10 μm JJs with 5 and 8 cycle ALD tunnel barrier is given in **Fig 4**. The 5 cycle sample was processed and measured in house, while the 8 cycle sample was processed and measured by collaborators at IOP, Beijing [16]. The supercurrent expected at zero voltage is missing from the IVC of the 5 cycle sample due to excessive noise in the measurement system; a standard sample was used to confirm that noise suppresses supercurrents and distorts the quasi-particle branch of the IVC. The supercurrent for the 8 cycle sample is also suppressed. We attribute this suppression to charged scattering centers (i.e. $OH^-$ groups and $H^+$ interstitials) either on the surface or of the ALD film or within it [16]. Both IVCs show a transition to the normal state near the gap voltage of Nb. $R_N$ is ~ 30 Ω and 50 Ω for the 5 and 8 cycle JJs respectively, which agrees reasonably with room temperature measurements. The sub-gap/normal-state resistance ratio is $R_{sg}/R_N \sim 2$ and 20 for 5 and 8 cycles respectively measured at V = 1.8 mV. We attribute this difference to better processing and measurement on the 8 cycle sample. The gap current $I_G$ = 100 μA and 60 μA for 5 and 8 cycles, respectively. In the absence of a supercurrent measurement, and since $I_G$ and $I_C$ are proportional [22], we use this to conclude that the critical current is controllable via ALD.

## IV. Conclusion

In conclusion, Nb-Al/ALD-$Al_2O_3$/Nb trilayers were fabricated *in situ*. The trilayers were patterned into JJs, which were characterized at room temperature and low temperature. A consistent $R_N A$ indicates the ALD tunnel barriers are uniform across the wafer, and an exponential increase in $R_N A$ with an increasing number of ALD cycles indicates the tunneling properties can be controlled by altering the number of ALD cycles performed. $R_N$ was extracted from room temperature IVCs, and an exponential increase in $R_N$ with increasing ALD cycles was observed. Further, extrapolation of this data to 0 ALD cycles indicates a thin $AlO_x$ IL was formed during ALD growth. Though this extrapolation does not agree with previous, low temperature measurements, AIMD simulations suggest the nucleation of ALD $AL_2O_3$ on Al is controlled by $H_2O$ during the first ALD cycle, and that substantial oxidation may occur that would account for the difference. Low temperature measurements corroborated the room temperature data. Together, these results indicate that ALD is a viable method for controlling the thickness of JJ tunnel barriers. However, an $AlO_x$ IL develops during the first few cycles of ALD growth, and it may be problematic for JJ qubits. Alternative wetting layer materials that are not susceptible to oxidation at 200 °C may be required to realize JJ qubits with ALD tunnel barriers.